\documentstyle[psfig]{mn}

\newif\ifAMStwofonts

\ifoldfss
  \ifCUPmtlplainloaded \else
    \NewTextAlphabet{textbfit} {cmbxti10} {}
    \NewTextAlphabet{textbfss} {cmssbx10} {}
    \NewMathAlphabet{mathbfit} {cmbxti10} {} 
    \NewMathAlphabet{mathbfss} {cmssbx10} {} 
  \fi
  \ifAMStwofonts
    \ifCUPmtlplainloaded \else
      \NewSymbolFont{upmath} {eurm10}
      \NewSymbolFont{AMSa} {msam10}
      \NewMathSymbol{\upi}     {0}{upmath}{19}
      \NewMathSymbol{\umu}     {0}{upmath}{16}
      \NewMathSymbol{\upartial}{0}{upmath}{40}
      \NewMathSymbol{\leqslant}{3}{AMSa}{36}
      \NewMathSymbol{\geqslant}{3}{AMSa}{3E}

    \fi
  \fi
\fi 

\ifnfssone
  \newmathalphabet{\mathit}
  \addtoversion{normal}{\mathit}{cmr}{m}{it}
  \addtoversion{bold}{\mathit}{cmr}{bx}{it}
  \newmathalphabet{\mathbfit} 
  \addtoversion{normal}{\mathbfit}{cmr}{bx}{it}
  \addtoversion{bold}{\mathbfit}{cmr}{bx}{it}
  \newmathalphabet{\mathbfss} 
  \addtoversion{normal}{\mathbfss}{cmss}{bx}{n}
  \addtoversion{bold}{\mathbfss}{cmss}{bx}{n}
  \ifAMStwofonts
    \ifCUPmtlplainloaded \else
      \UseAMStwoboldmath
      \makeatletter
      \new@mathgroup\upmath@group
      \define@mathgroup\mv@normal\upmath@group{eur}{m}{n}
      \define@mathgroup\mv@bold\upmath@group{eur}{b}{n}
      \edef\UPM{\hexnumber\upmath@group}
      \new@mathgroup\amsa@group
      \define@mathgroup\mv@normal\amsa@group{msa}{m}{n}
      \define@mathgroup\mv@bold\amsa@group{msa}{m}{n}
      \edef\AMSa{\hexnumber\amsa@group}
      \makeatother
      \mathchardef\upi="0\UPM19
      \mathchardef\umu="0\UPM16
      \mathchardef\upartial="0\UPM40
      \mathchardef\leqslant="3\AMSa36
      \mathchardef\geqslant="3\AMSa3E
    \fi
  \fi
\fi 

\ifnfsstwo
  \DeclareMathAlphabet{\mathbfit}{OT1}{cmr}{bx}{it}
  \SetMathAlphabet\mathbfit{bold}{OT1}{cmr}{bx}{it}
  \DeclareMathAlphabet{\mathbfss}{OT1}{cmss}{bx}{n}
  \SetMathAlphabet\mathbfss{bold}{OT1}{cmss}{bx}{n}
  \ifAMStwofonts
    \ifCUPmtlplainloaded \else
      \DeclareSymbolFont{UPM}{U}{eur}{m}{n}
      \SetSymbolFont{UPM}{bold}{U}{eur}{b}{n}
      \DeclareSymbolFont{AMSa}{U}{msa}{m}{n}
      \DeclareMathSymbol{\upi}{0}{UPM}{"19}
      \DeclareMathSymbol{\umu}{0}{UPM}{"16}
      \DeclareMathSymbol{\upartial}{0}{UPM}{"40}
      \DeclareMathSymbol{\leqslant}{3}{AMSa}{"36}
      \DeclareMathSymbol{\geqslant}{3}{AMSa}{"3E}
    \fi
  \fi
\fi 

\ifCUPmtlplainloaded \else
  \ifAMStwofonts \else 
    \def\upi{\pi}
    \def\umu{\mu}
    \def\upartial{\partial}
  \fi
\fi

\title{High-resolution single-pulse studies of the Vela Pulsar}
\author[Kramer, Johnston \& van Straten]{ M.~Kramer,$^1$
S.~Johnston,$^{2}$ W. van Straten$^3$ \\
$^1$University of Manchester, Jodrell Bank Observatory, Macclesfield,
Cheshire SK11 9DL, UK \\
$^2$School of Physics, University of Sydney, NSW 2006, Australia \\
$^3$Swinburne Centre for Astrophysics and Supercomputing, Swinburne 
University of Technology, Hawthorn, VIC 3122, Australia}

\date{10 February 2002}

\pagerange{\pageref{firstpage}--\pageref{lastpage}}
\pubyear{2002}

\begin{document}

\maketitle

\label{firstpage}

\begin{abstract}
We present high-resolution multi-frequency single-pulse observations of the
Vela pulsar, PSR B0833$-$45, aimed at studying micro-structure,
phase-resolved intensity fluctuations and energy distributions at  1.41
and 2.30 GHz. We find micro-structure in about 80\% of all
pulses independent of observing frequency.  The width of a
micro-pulse seems to depend on its peak flux density, whilst
quasi-periodicities observed in micro-pulses remain constant.  The
width of the micro-pulses may decrease with increasing frequency, but
confirmation is needed at higher frequencies. The fraction of pulses
showing quasi-periodic micro-pulses may become smaller at higher
frequencies.
We show that the micro-pulse width in pulsars has a
period dependence which suggests a model of
sweeping micro-beams as the origin of the micro-pulses.

Like individual pulses, Vela's micro-pulses are highly elliptically
polarized.  There is a strong correlation between Stokes parameters
$V$ and $I$ in the micro-structure. We do not observe any micro-pulses
where the handedness of the circular intensity changes within a
micro-pulse although flips within a pulse are not uncommon.  We show
that the $V$/$I$ distribution is Gaussian with a narrow width and that
this width appears to be constant as a function of pulse phase.  The
phase-resolved intensity distributions of $I$ are best fitted with
log-normal statistics.

Extra emission components, i.e.~``bump'' and ``giant 
micro-pulses'', discovered by Johnston
et al.~(2001) at 0.6 and 1.4 GHz 
are also present at the higher frequency of 2.3 GHz. 
The bump component seems to be an
extra component superposed on the main pulse profile but does not
appear periodically. The giant micro-pulses are time-resolved and have
significant jitter in their arrival times. Their flux 
density distribution is
best fitted by a power-law, indicating a link between these
features and ``classical'' giant pulses as observed for the Crab pulsar,
(PSR B0531+21), PSR B1937+21 and PSR B1821$-$24.

We find that Vela contains a mixture of emission properties representing
both ``classical'' properties of radio pulsars
(e.g.~micro-structure, high degree of polarization, 
S-like position angle swing, orthogonal modes) and features which are
most likely related to high-energy emission (e.g.~extra profile components,
giant micro-pulses). It hence represents an ideal test case to study the
relationship between radio and high-energy emission in significant detail.
\end{abstract}


\section{Introduction}
\label{intro}

After more than thirty years of pulsar observations,
the emission mechanism of pulsars is only poorly
understood. One of the reasons for the absence of 
an accepted radiation
theory is the variety of pulse phenomena, acting on several different
time scales, which have to be explained. Typical time scales are
present in the following forms: the rotation period of the neutron
star, $P$, and the typical width of sub-pulses constituting an
individual pulse of emission, $\tau_s$ \cite{dc68}.  For a number of
pulsars, these sub-pulses drift across the pulse window with a
certain drift rate that is determined by the separation of two
drifting sub-pulses within a single pulse and the period at
which sub-pulses re-occur at the same pulse phase
\cite{bac70c}. Yet, there is another sub-class of pulsars, for which
{\em micro-structure} has been identified \cite{ccd68}. Usually sitting
on top of sub-pulses, {\em micro-pulses} appear to be concentrated
features of emission which often exhibit typical widths $\tau_\mu$ and
sometimes appear quasi-periodic within a sub-pulse, showing a
periodicity, $P_\mu$ \cite{han71}.

A fundamental problem in the interpretation of these features seen
in pulsar emission is the determination
which of the features are truely representative of the
radiation process and which are caused by propagation effects in the
magnetosphere. At least for micro-structure the answer seems to be
clear. It was shown that micro-pulses appear simultaneously at
different, widely spaced frequencies (when corrected for dispersion,
e.g. Rickett, Hankins \& Cordes 1975; Boriakoff, Ferguson \& Slater 
1981\nocite{rhc75,bfs81}) which would indicate a fundamental association
with the emission process.  However, it still remains to be shown 
whether micro-structure is a {\em radial} or a 
{\em temporal} effect. In the first
case, micro-pulses may be due to individual beams similar to sub-pulse
beams sweeping across the line-of-sight. 
In the latter, temporal modulations within a plasma column or
retardation delays over a radial range may be responsible 
(e.g.~Cordes 1979\nocite{cor79b}).

The detection of micro-pulses requires a sufficiently high time
resolution during the observations. While at higher frequencies the
dispersion smearing is greatly reduced for a given bandwidth, 
only coherent de-dispersion
techniques have usually been powerful enough at lower frequencies
(e.g. Hankins 1971\nocite{han71}).
Another limiting factor is the additional effect
of interstellar scattering, which can smear the pulses sufficiently to
prevent the detection of micro-structure. A combination of both effects
is likely to be 
responsible for the fact that there has not been any published
micro-structure study for the brightest pulsar in the sky, the Vela
pulsar, PSR B0833$-$45. Until recently, the only
detailed single-pulse study of Vela to have appeared in the 
literature is that by Krishnamohan \& Downs (1983; hereafter
KD83\nocite{kd83}).  For their study, KD83 observed 87000 pulses at
2.3 GHz with a time resolution (including dispersion
smearing) of only 750$\mu$s, and make no mention of micro-structure.
Recently, however, in the first
part of this work (Johnston et al.~2001, hereafter Paper I)
\nocite{jvkb01} we presented high time resolution
observations of single pulses from the Vela pulsar made with a
baseband recording system at observing frequencies of 650 and 1413
MHz.  A large fraction of the pulses exhibit micro-structure, which
will be the subject of this paper.

In Paper I, we noted the presence of large amplitude pulses which
occur before the start of the main pulse. We called these ``giant
micro-pulses'' as they are often 100$\times$ the expected flux density
{\em at those pulse phases} but have only rather little effect on the
pulse integrated flux density.  This is in contrast to the
``classical'' giant pulses seen in the Crab and the millisecond
pulsars PSR B1937+21 and B1821$-$24, where the flux density in a
single pulse can be more than 10$\times$ the {\em mean} flux
density. It seems likely that these giant pulses are related more to
the high-energy emission from these pulsars than to the conventional
radio emission \cite{rj01}.  In this paper we explore the relationship
between the giant micro-pulses and the rest of the radio emission in
more detail.

\begin{figure*}

\centerline{\psfig{figure=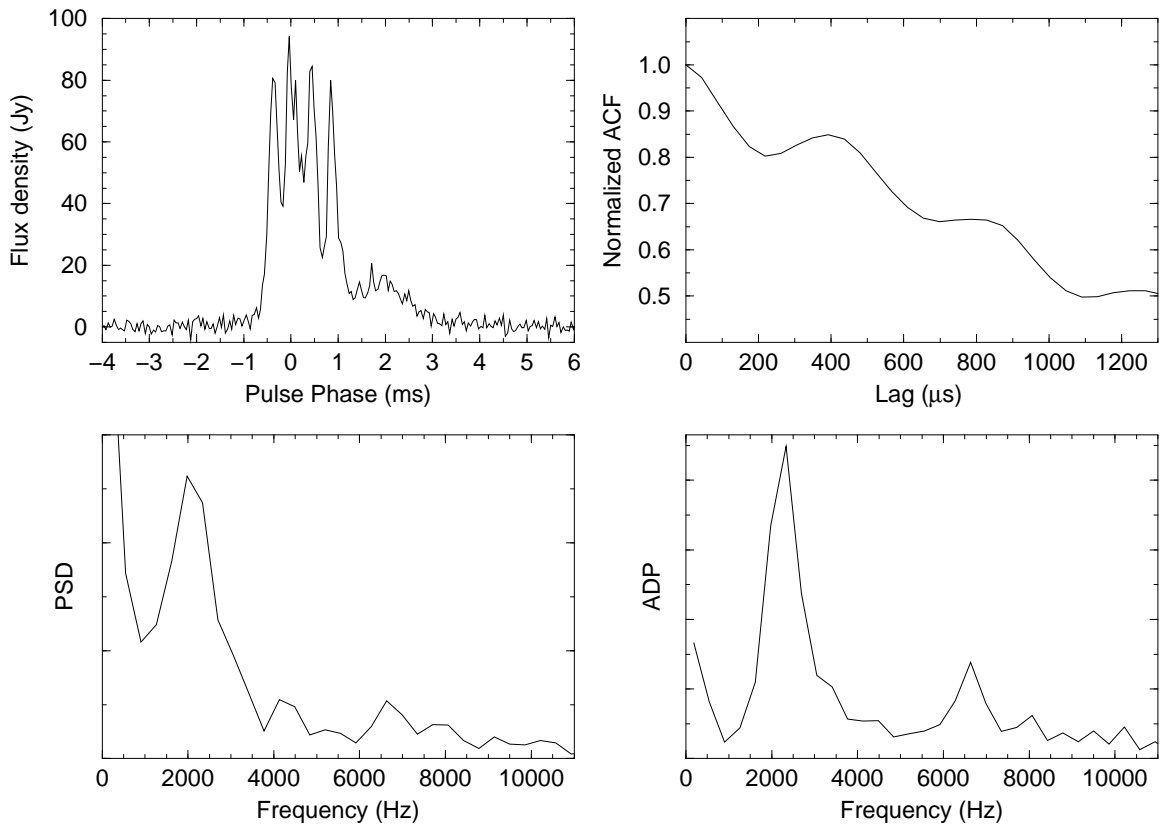,height=12cm,angle=-90}}

\caption{\label{pulseexp}
Example of a typical pulse at 1.4 GHz showing micro-structure (top left) 
and its computed ACF (top right), PSD (bottom left) and ADP
(bottom right). The quasi-periodicity in the micro-pulses is visible
in the ACF, the PSD and in particular in the ADP.}
\end{figure*}

\section{Observations\label{obs}}
We made observations of Vela
on 2000 March 15 and 2001 March 28/29
using the 64-m Parkes radio-telescope.  The centre frequencies of the
observations were 1.413 and 2.295 GHz; at these
frequencies the system equivalent flux densities are 26 and 40 Jy
respectively.  Each receiver consists of two orthogonal feeds sensitive to
linear polarization at 1.4~GHz and circular polarization at 2.3~GHz. The
signals are down-converted and amplified before being passed into the backend.
The backend, CPSR, is an enhanced version of the Caltech Baseband Recorder
\cite{jcpu97}. It consists of an analogue dual-channel quadrature converter
which samples data with 2-bit precision at 20 MHz. The data stream is written
to DLT for subsequent off-line processing.  At 1.4 GHz, $30$ minutes of
data or $\sim$20000 pulses were recorded and analysed in each run.  At 2.3
GHz we recorded $\sim 120000$ pulses or 3 hours of continuous data.  Before
each observation, a 90-second observation of a pulsed signal, directly
injected into the receiver at a $45\degr$ angle to the feed element, is made.
This enables instrumental polarization to be corrected. Observations of the
flux density calibrator Hydra A were made at each of the observing
frequencies; this allows absolute flux densities to be obtained.

The data were calibrated and processed off-line using a 
workstation cluster at the Swinburne Supercomputer Centre using the 
technique described in Paper I.  At 2.3 GHz
there are 4096 time-bins per pulse period, yielding an effective resolution of
21.8 $\mu$s, which is considerably less than the scatter broadening.  At 1.4
GHz there are 2048 for a time resolution of 44 $\mu$s, corresponding to about
twice the time-scale of the interstellar scattering. Table~\ref{resulttab}
lists the relevant observing details.

\begin{table*}
\caption{\label{resulttab}
Summary of observations and determined micro-structure parameters.
We list the scattering time computed for each frequency, the number
of pulses observed, the fraction of pulses showing micro-structure,
the typical micro-pulse width, the fraction of pulses show 
quasi-periodic micro-structure and corresponding periodicities
as determined from ACFs, PSDs and ADPs.}
\begin{tabular}{ccccccccc}
\hline
\hline
Epoch & Frequency & $\tau_{\rm scatt}$ & Pulses & Fraction$_{\tau_\mu}$ 
&  $\tau_\mu$ 
& Fraction$_{P_\mu}$ &   $P_\mu^{ACF}$ & $P_\mu^{PSD/ADP}$ \\
 & (GHz) & ($\mu$s) &  &  (\%) & ($\mu$s) & (\%) & ($\mu$s) & ($\mu$s) \\
\hline
March 2000 & 1.41 & 21 & 20085 & 82 &  $220\pm80$ &  35 & $450\pm260$ & 
$470\pm160$ \\
March 2001 & 1.41 & 21 & 23999 & 78 &  $220\pm90$ &  32 & $610\pm260$ & 
$470\pm160$ \\
March 2001 & 2.30 & 3 & 122280 & 85 &  $130\pm40$ &  25 & $410\pm280$ & 
$470\pm80$ \\
\hline
\end{tabular}
\end{table*}

\section{Micro-structure}
\label{micro}

\subsection{Data analysis}

\label{micromethod}

When studying the micro-structure properties of individual
(sub-)pulses we are interested in identifying typical micro-pulse
widths, $\tau_\mu$, and the possible existence of (quasi-)
periodicities. The standard analysis procedure is the computation of
autocorrelation functions (ACFs). In order to determine
quasi-periodicities, a power spectrum of the data (PSD) can be
calculated, or alternatively the power spectrum of the ACF derivative,
the so-called ADP.  The latter, introduced by Lange et al.~(1998),
produces similar results as a PSD but performs a frequency weighting
without adding spurious frequencies.  In the present analysis we
compute ACFs, PSDs and ADPs for all pulses and compute their average
for the whole ensemble of pulses. In the following we give a brief
overview over the analysis techniques; interested readers should refer
to Lange et al.~(1998)\nocite{lkwj98} for details.

For each pulse, a sub-pulse window 64 time-bins wide at 1.4 GHz and a
window of 128 time-bins wide at 2.3 GHz is automatically determined for each
pulsar by finding the maximum of a cross-correlation of the individual pulse
with a box-car window of same length. For each window covering 
data of length 2.792 ms the ACF is
computed. If a micro-pulse is present, the ACF shows a change in slope at the
lag corresponding to $\tau_\mu$ (the first order turn-off point).  A further
slope change indicates the typical sub-pulse width $\tau_s$ (second order
turn-off point).  A quasi-periodicity existing within the pulse is noticeable
as equally spaced peaks in the ACF, and the lag of the first peak indicates
the periodicity $P_\mu$.  Additionally, PSDs and ADPs are computed and peaks
recorded along with the values found from the ACF analysis.  In
Figure~\ref{pulseexp} we show a typical example of the micro-structure analysis
on a single pulse at 1.4 GHz.

\subsection{Results}

\begin{figure*}

\begin{tabular}{ll}
a) & b) \\
\psfig{figure=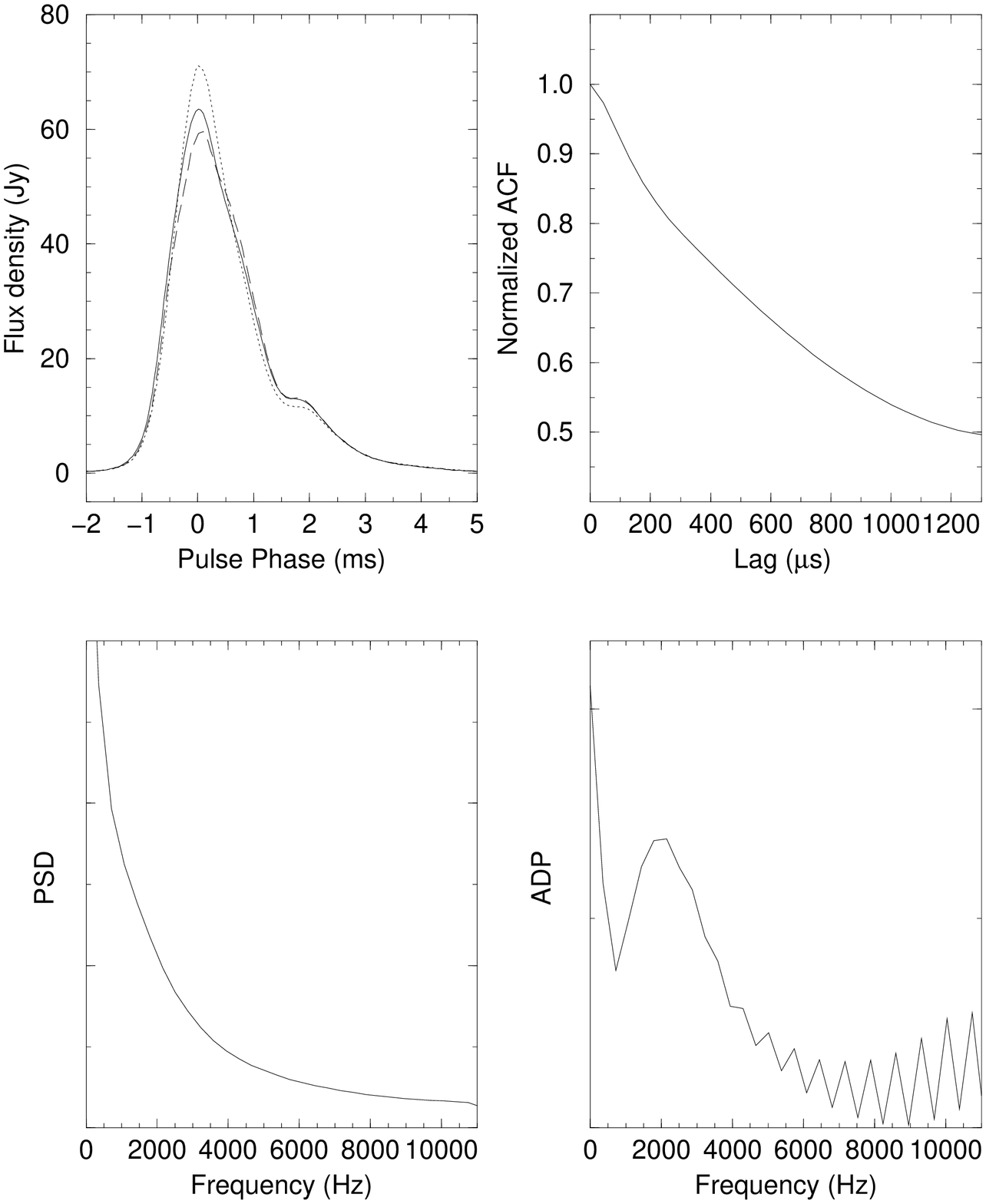,height=12cm}
&
\psfig{figure=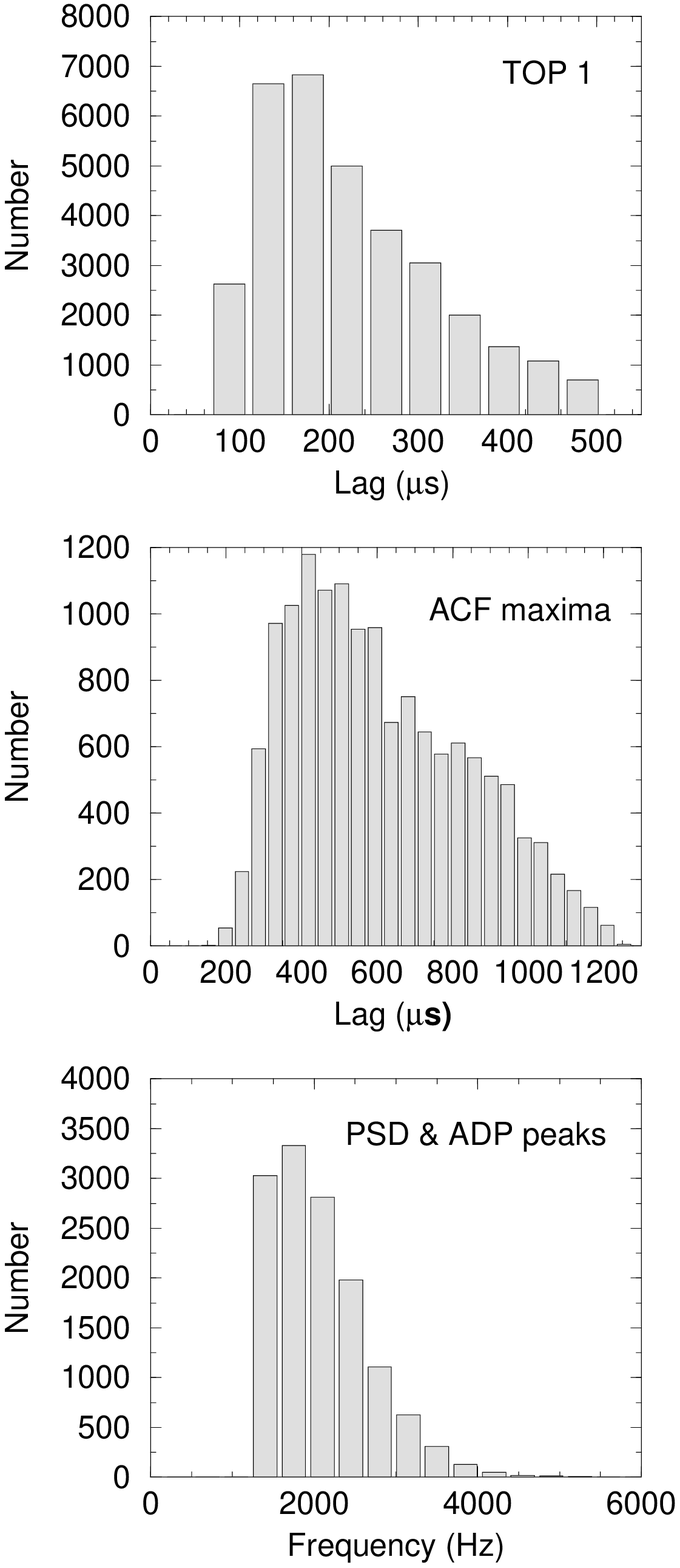,height=12cm}
\end{tabular}

\caption{\label{resultfiga}\label{resultfigb}
Results of the micro-structure analysis for 1.4 GHz. 
a) ({\sl top}) Average pulse profile(s),
average normalized ACF, ({\sl bottom}) average PSD and average ADP, b)
distributions derived from individual pulses as first order turn-off
points in the ACF, first maxima in the ACF and first significant peaks
simultaneously occurring in the PSD and ADP. For the mean pulse
profiles in a) we show the average of all pulses (solid), the average
of pulses exhibiting micro-structure (dashed) and the average of pulses
without micro-structure (dotted).
}
\end{figure*}

The results of the micro-structure analysis are presented both in the form
of distributions of all relevant values obtained from the observed
pulses, as well as in the mean ACF, PSD and ADP, averaged over all
pulses at a given frequency. The determined quantities are summarized
in Figures~\ref{resultfiga} and 
in Table~\ref{resulttab} as medians and
uncertainties as determined from the histograms. Notice that the
various methods yield consistent results, e.g.~quasi-periodicities can
be determined from the peak in the average ADP
(Fig.~\ref{resultfiga}a, bottom right), the distribution of the
maxima in the individual ACFs (Fig.~\ref{resultfigb}b, middle), and 
the distributions of maxima found simultaneously in individual PSDs
and ADPs (Fig.~\ref{resultfigb}b, bottom).

At both frequencies we consistently detect micro-structure features in
about 80\% of all pulses. Hence, our results do not indicate that the
micro-pulse component has a steeper spectrum as, for instance,
inferred for PSR B2016+28 by Cordes et al.~(1990). This is consistent
with the finding of Lange et al.~(1998) who detected micro-structure
in virtually every pulsar strong enough to be observed at 4.85 GHz.
We believe that previous reports on steeper spectra can be attributed
to a limited signal-to-noise ratio available in those studies. Therefore,
micro-structure appears to be a fundamental integral feature
of pulsar radio emission.

Interestingly, we appear to detect a
typical micro-structure time scale which is shorter at 2.3 GHz than at
1.4 GHz although the statistical significance is not very high. 
In order to test whether this is a result of the higher time
resolution, we have performed our analysis on 2.3 GHz data modified to
match the resolution at 1.4 GHz. We obtain the same result again,
i.e.~a smaller time scale at the higher frequency. In contrast, the
determined periodicities are clearly identical within the errors and also
agree among the different methods applied to obtain them.  Hence, the
micro-pulses may become narrower at 2.3 GHz but keep the same spacing if
they appear in a quasi-periodic fashion. The fraction of
micro-pulses for which we detect quasi-periodicities seems to be
slightly smaller at the higher frequency. However, we cannot fully rule out
a difference caused by a smaller signal-to-noise ratio at 2.3 GHz, or
alternatively, by an effect described in the following.

\begin{figure}

\centerline{\psfig{figure=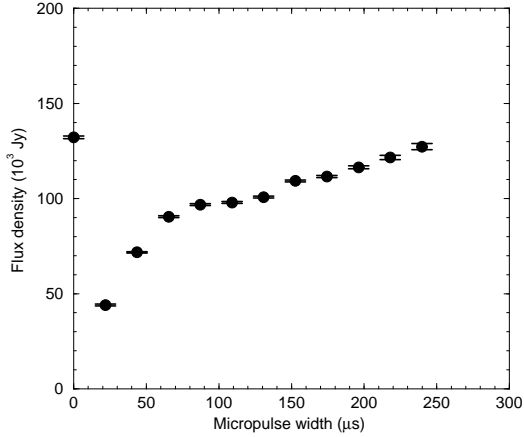,height=6cm}}

\caption{\label{widthflux}
Mean peak flux density as function of micro-pulse time scale as
determined at 2.3 GHz. Peak flux densities were recorded for each
analysis window along with determined time-scales. A width of zero
indicates that no micro-pulse time scale was detected in the ACF.}
\end{figure}

In Figure \ref{widthflux} we show the average peak flux density of a
micro-pulse at 2.3 GHz as a function of its width measured from the
first turn-off point in the ACF.  Those pulses for which we cannot
identify a turn-off point and, hence, which are not identified to
contain a micro-pulse of a certain width, are collected in the zero-th
bin of this figure. This obviously happens typically for a micro-pulse
width exceeding 250$\mu$s, which is identical to half of the
quasi-periodicity (see Table~\ref{resulttab}). At this point, the
micro-pulses obviously blend into each other and are not identified as
separate entities any longer. The result of this effect and the
increased peak flux density with larger micro-pulse width is the
formation of significantly different pulse profiles when computed from
those pulses for which we detect micro-structure and from those where
we do not (see Fig.~\ref{resultfiga}). While this effect confirms the
determined average quasi-periodicity, it remains to be explained what
causes the apparent correlation between micro-pulse peak flux density
and width. 
In any case this interpretation suggests that the fraction
of pulses containing micro-structure is even larger than the 80\%
we derive.

\subsection{A temporal or radial origin of micro-structure?}

Table \ref{previousobs} lists the pulsars which are known to have
micro-structure. 
%
%
Comparing the micro-pulse separations, $P_\mu$, for different pulsars
and frequencies, they seem to be frequency independent as previously
summarized by Cordes et al.~(1990). This is also true for the
periodicities found for Vela at 1.4 GHz and 2.3 GHz as given in
Table~\ref{resulttab}. The situation is, however, less clear for the
micro-structure time-scale.  Our Vela results may hint that
micro-pulses may become narrower at higher frequencies, although this
is not statistically significant and higher frequency data are
needed. For other pulsars listed in Table~\ref{resulttab} the
uncertainties are often large -- if multi-frequency data exist at all.
The best studied case is PSR B1133+16, and there is indeed some indication
that its time-scale is decreasing at higher frequencies: Hankins
et al.~(1976\nocite{hcr76}) reported a dependence of $\tau_\mu\propto
\nu^{-0.22\pm0.05}$ which is supported by the results of Ferguson \&
Seiradakis (1978). However, the compilation of PSR B1133+16 data in 
Table~\ref{resulttab} makes these results less obvious, yielding a formal fit of
$\tau_\mu\propto\nu^{-0.06\pm0.10}$.

\begin{table}
\caption{\label{previousobs}
Summary of previous micro-structure observations.}
\begin{tabular}{lccccl}
\hline
\hline
PSR B & Period & Freq. & $\tau_\mu$ & $P_\mu$ & Ref. \\
    & (s)    &   (GHz)   &  ($\mu$s)  & ($\mu$s) &           \\
\hline
0329+54 & 0.715 & 0.10   &  ...    & 340        & KKN78 \\
        &       & 4.85   & $<1500$ & $600-1500$ & LKWJ  \\
0525+21 & 3.746 & 0.43   & 3000    &   ...      & COR79 \\  
0540+23 & 0.246 & 1.41   & $<360$  &   ...      & LKWJ \\
0809+74 & 1.292 & 0.10   & 770     &  5000      & CWH90 \\
0823+26 & 0.531 & 0.43   & 550     &   ...      & COR79 \\
        &       & 4.85   & $300-700$ & $360-660$  & LKWJ  \\
0834+06 & 1.273 & 0.43   & 1050    &   ...      & COR79 \\
0950+08 & 0.253 & 0.11   & 100 & 400         & CWH90 \\
        &       & 0.11   & 200 &    ...      & RHC75 \\
        &       & 0.32   & 200 &    ...      & RHC75 \\
        &       & 0.43   & 100 & 400         & CWH90 \\
        &       & 0.43      & 170 &    ...      & CH77 \\
        &       & 1.65      & 140 &    ...      & PBC01 \\
        &       & 4.85      & 170 &    ...      & LKWJ \\
1133+16 & 1.188 & 0.10      & 500 &    ...      & PSS87 \\
        &       & 0.10      & 500 &    ...      & PSS87 \\
        &       & 0.10      & 574 &    ...      & HAN72 \\
        &       & 0.20      & 651 &    ...      & HAN72 \\
        &       & 0.32      & 525 &    ...      & HAN72 \\
        &       & 0.43      & 340 &    ...      & CH77 \\
        &       & 0.43      & $130-200$ & 800   & CWH90 \\
        &       & 0.61       & $130-200$ & 800   & CWH90 \\
        &       & 1.42      & 429 &    ...       & FGJ76 \\
        &       & 1.65      & $<10/300$ & 800/400 & PBC01 \\
        &       & 1.72      & 380  & $180-1090$   & FS78 \\
        &       & 2.65      & 420  & $180-1090$   & FS78 \\
        &       & 4.85      & 365  & $<800$      & LKWJ \\
1919+21 & 1.337 & 0.43      & 1300 & ...        & COR79 \\
1929+10 & 0.227 & 1.65      & 90   & 300        & PBC01 \\
        &       & 4.85      & 150  & $240-1000$ & LKWJ \\
1944+17 & 0.441 & 0.43      & 300  & ...        & COR79 \\
        &       & 0.43      & 160  & 800        & CWH90 \\
2016+28 & 0.558 & 0.32      & 160  & 800        & CWH90 \\
        &       & 0.43      & 290  & 900        & COR76 \\
        &       & 0.43      & 160  & 800        & CWH90 \\
        &       & 0.61      & ...  & $500-750$  & COR76 \\
        &       & 0.61      & 160  & 800        & CWH90 \\
        &       & 1.41      & 160  & 800        & CWH90 \\
        &       & 1.41      & 240  & 640        & LKWJ \\
2020+28 & 0.343 & 0.43      & 110  & ...        & COR79 \\
\hline
\end{tabular}
{\footnotesize
References: 
HAN72 - Hankins 1972,\nocite{han72}
RHC75 - Rickett et al.~1975\nocite{rhc75}
COR76 - Cordes 1976,\nocite{cor76}
CH77 - Cordes \& Hankins 1977,\nocite{ch77}
FGJ76 - Ferguson et al.~1976, \nocite{fgj+76}
FS78 - Ferguson \& Seiradakis 1978,\nocite{fs78}
KKN78 - Kardashev et al.~1978, \nocite{kkn+78}
COR79 - Cordes 1979, \nocite{cor79b}
PSS87 - Popov et al.~1987,\nocite{pss87}
CWH90 - Cordes et al.~1990,\nocite{cwh90}
LKWJ - Lange et al.~1998,\nocite{lkwj98}
PBC01 - Popov et al.~2001\nocite{pbc+01}
}
\end{table}

\begin{figure}

\centerline{\psfig{figure=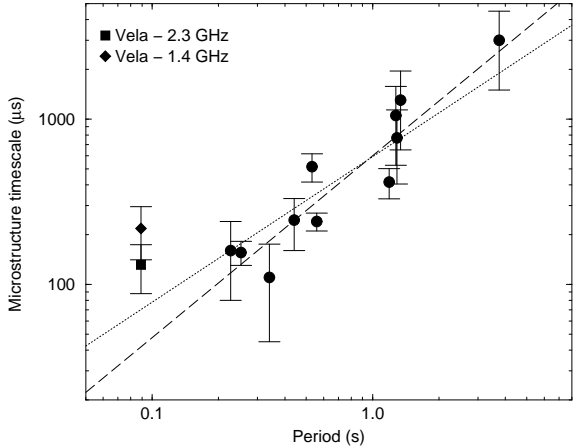,height=6cm}}

\caption{\label{allpsrs}
Micro-structure time-scales derived for Vela at 1.4 GHz and 2.3 GHz, and
time-scales reported for other pulsars as function of pulse period.
The dashed line represents a best fit to all data but those determined
for Vela. The dotted line is a fit including an averaged micro-pulse width for
Vela.
}
\end{figure}

Figure \ref{allpsrs} shows micro-pulse widths for different pulsars as a 
function of pulse period, $P$. We emphasize that this figures combines
results derived at different frequencies (or averages of multi-frequency 
data if available for a given pulsar). This may introduce some scatter if 
a change in micro-pulse width with frequency is confirmed, but it is necessary
to improve the otherwise small statistics. A similar plot using a smaller
number of sources was presented by Cordes (1979) who
suggested a linear dependence of $\tau_\mu$ on $P$. Obviously, this
relationship still holds.  The dashed line fitted to the data in
Figure~\ref{allpsrs} corresponds to
\begin{equation}
\label{scaling}
\tau_\mu \; \mbox{($\mu$s)} = (600\pm100) \; P\mbox{(s)}^{1.1\pm0.2}
\end{equation}
Interpreting such a linear dependence on period, one should be
aware that the time resolution during observations usually scales
similar, suggesting a possible instrumental origin of this result.  We
can reject such a possibility for our observations as we have analysed
our data using different resolutions achieved by re-binning the
data. Moreover, Popov et al.~(2001) have studied micro-structure of
PSR B1133+16 at 1.65 GHz with a resolution of only 62.5 ns. Although
they find some features to be as short as 10$\mu$s, the vast majority
of their detected micro-pulses show widths that are consistent which
previous low resolution results, supporting a physical meaning of the
above period trend.

In Fig.~\ref{allpsrs} we also show the time-scales determined for
Vela's micro-structure at 1.4 GHz and 2.3 GHz, respectively, which
were not included in the fit leading to Eqn.~(\ref{scaling}).
Including the data in the fit results in a
$P^{0.9\pm0.1}$--dependence (dotted line), 
i.e.~making the slope only marginally
flatter. Due to the typically large uncertainties, this fit is
statistically not very different from Eqn.~(\ref{scaling})
(i.e.~$\chi^2=3.4$ vs. $\chi^2=2.5$).  While it would have interesting
implications for the interpretation of micro-structure if the results
for Vela  indicated a flattening of the
period relationship towards lower periods, such a conclusion is not
justified given the current data. Observations of even faster rotating
pulsars are required. In any case, however, it seems to be established
that there is a clear period dependence of the width of normal
micro-pulses, suggesting a temporal origin of micro-structure.

Finally, we note that giant micro-pulses are typically narrower than
the normal ones, e.g.~measuring the width (FWHM) of five giant
micro-pulses with the largest S/N observed at 2.3 GHz, we obtain
widths of 114 $\mu$s (338 Jy), 73 $\mu$s (213 Jy), 100 $\mu$s (100
Jy), 38 $\mu$s (232 Jy), 41 $\mu$s (172 Jy) with a typical uncertainty
of 21 $\mu$s.  Such widths would fit even better to
Eqn.~(\ref{scaling}), which predicts a time-scale of about
$40^{+40}_{-20}$ $\mu$s for Vela.  However, we will argure later that
giant micro-pulses may have a different origin to the normal
micro-pulsars as, for example, we do not find any correlation between
the strength and the width of giant micro-pulses.

\label{giant1}

\section{Fluctuation analysis}

A fluctuation analysis was carried out for the Vela pulsar by
KD83. Their observations made at a frequency of 2.3 GHz was limited to
an effective time resolution of 750 $\mu$s, i.e.~a factor of 34 worse
than our resolution at the same frequency. Two spectral features
located at non-zero frequencies were found by KD83. In the trailing part
of the pulse profile they identified a feature at 0.031 cycles
period$^{-1}$ and in the right edge of the main pulse peak a broad
spectral feature at 0.125 cycles period$^{-1}$.

\begin{figure*}
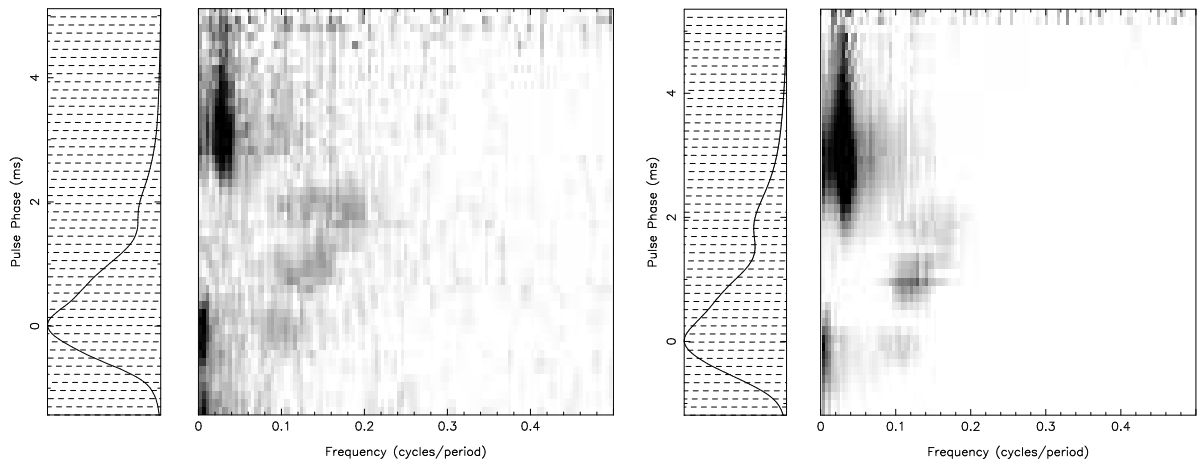


\begin{center}
\begin{tabular}{cc}
\psfig{figure=fig5a.ps,height=6cm,angle=-90}
&
\psfig{figure=fig5b.ps,height=6cm,angle=-90} 
\end{tabular}
\end{center}

\caption{\label{flucfig}
Average pulse profiles and fluctuation spectra computed from 12672 pulses
at 1.4 GHz (left) and from 114688 pulses at 2.3 GHz (right).
Four significant features at non-zero frequency are clearly visible.}
\end{figure*}

Using our high resolution data, we have also performed a fluctuation
spectra analysis in which we essentially follow the example of Backer
(1970\nocite{bac70c}, 1973\nocite{bac73}).  Using continuous sets of
individual pulses, we compute PSDs for the time series of pulse energy
present in 50 adjacent windows covering the pulse profile in bins of
123 $\mu$s width each.  Additionally, we calculate the PSD for the
time series of an off-pulse window of the same size, which is
subtracted from each of the on-pulse window PSDs.  Usually, the power
in each spectrum is normalized to match the power of the corresponding
time series. However, in order to search for signals in weaker parts
of the pulse profiles, we alternatively normalize the spectrum to a
uniform RMS.  In order to increase the signal-noise-ratio of spectral
features at the cost of spectral resolution, we also compute the
incoherent sum of power spectra obtained from Fast Fourier Transforms
of smaller parts of the total available time series. Results are
presented for 1.4 GHz and 2.3 GHz in Fig.~\ref{flucfig}.  We find the
same spectral features at both frequencies with a high degree of
significance.  Values for the pulse phase range, the determined
centroid of the frequency feature and a frequency width estimated for
a 10\% level are listed in Table~\ref{fluctab}. Note that the features
apparently existing at very small frequencies at phases $-$1.29 to
0.23 ms and 1.84 to 3.81 ms are consistent with zero frequency,
similar to findings of KD83.  Features 2 and 4 are obviously identical
to those observed by KD83. Due to our better time resolution and
sensitivity, we can however separate KD83's second feature into three,
apparently distinct features, here called 1, 2 and 3 (note the drop in
power around these ``frequency islands'' in Fig.~\ref{flucfig}).  We
note that when comparing the determined centroids of these features,
they seem to lie on a straight line with a slope of $0.025\pm0.005$
(cycles period$^{-1}$) ms$^{-1}$. There are some indications that
feature 2 is also visible in the longitude range of 1.82 to 1.93 ms,
partly overlapping with feature 3.  We performed an analysis as
described by Deshpande \& Rankin (2001\nocite{dr01}) which shows that
all three features represent the fundamental frequencies, i.e.~they do
not originate from higher frequencies being aliased due to the finite
sampling once a period.

\begin{table}
\caption{\label{fluctab}
Spectral features derived from the fluctuation analysis. We quote
the centre of the spectral feature, as well as their estimated 
width measured at a 10\% level.}
\centerline{
\begin{tabular}{lr@{~--~}rccr}
\hline
\hline
 & \multicolumn{2}{c}{Pulse Phase} & Frequency & Width & S/N \\
   & \multicolumn{2}{c}{(ms)} & (cycles period$^{-1}$) & (cycles period$^{-1}$) & \\
\hline
1 & $-0.30$ & 0.14 & $0.093\pm0.002$  & $0.019\pm0.002$ & 4  \\
2 & 0.68    & 1.13 & $0.125\pm0.003$  & $0.031\pm0.003$ & 36  \\
3 & 1.57    & 1.93 & $0.134\pm0.003$  & $0.041\pm0.004$ & 8  \\
4 & 1.48    & 4.52 & $0.0335\pm0.0004$ & $0.0018\pm0.0001$ & 68  \\
\hline
\end{tabular}
}
\end{table}

In Paper I we reported an extra profile feature,
occurring sometimes as a strong ``bump'' component at the edge of the
trailing profile, and speculated as to whether its appearance was periodic.
Using our extremely long observation at 2.3 GHz, we can now 
rule out any significant periodicity for the occurrence of
the bump, although the broad feature No.~4 listed in
Table~\ref{fluctab} partly covers this region.


\section{Pulse Energy Distribution}

\label{distsect}

\begin{figure}
\centerline{\psfig{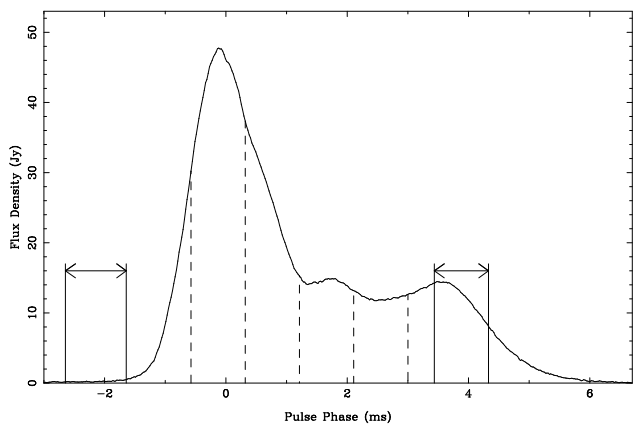}}
\caption{\label{bumpfig}
The integrated profile at 2.3 GHz of 2355 single pulses with
flux densities greater than 5$\times$ the mean in the bump. The 
bump is visible as the trailing resolved peak of the profile.
Dashed lines refer to pulse phases referred to in the text.
The solid lines on the left denote the giant micro-pulse region, the
pair on the right denote the bump region used in the analysis
described in the text.}
\end{figure}

The recent literature on the energy distribution of single pulses is
rather sparse (except for those pulsars showing giant pulses). In
Paper I we showed that the logarithm of the flux density of single
pulses from the Vela pulsar had a Gaussian distribution (a so-called
log-normal distribution).  In retrospect, the distributions of single
pulse energies shown in Ritchings (1976)\nocite{rit76} and Hesse \&
Wielebinski (1974)\nocite{hw74} for a variety of pulsars also appear
to be log-normal although they are not discussed in that way by those
authors.  Cairns et al. (2001)\nocite{cjd01} showed that the
phase-resolved intensity distributions also followed log-normal
statistics in the Vela pulsar at 1.4 GHz. In this section we
investigate different pulse phases at 2.3 GHz and examine their flux
density 
distributions. Figure \ref{bumpfig} shows the integrated pulse profile
at 2.3 GHz, where we have picked out the strongest $\sim$2000 pulses
showing the ``bump'' feature. Vertical lines mark the pulse phases of
interest.

\subsection{Main pulse window}

\begin{figure}
\centerline{\psfig{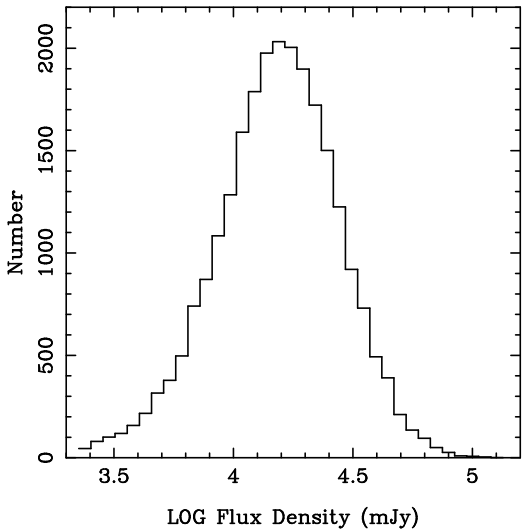}}
\caption{\label{culmfig2}
Histogram of energy distributions for phase 1.213 ms.}
\end{figure}

We examined five phases (bins centred on the 
dashed lines in Fig.~\ref{bumpfig}) equally
spaced across the main pulse window.  All of the flux density
distribution of these bins can be fit with a log-normal distribution
of pulse intensities convolved with a Gaussian noise distribution with
constant mean 0.0 and sigma 2250 mJy.  Table \ref{culmtab} summarises the
means and sigmas of the fitted log-normal distributions. 
For illustration, Figure \ref{culmfig2} shows
the flux density distribution and cumulative probability distribution
for phase 1.213 ms. 

It is evident from the table that on the rising
part of the pulse profile, the width of the distribution is
large. This is the equivalent to saying that the modulation index in
this part of the pulse is larger than that towards the middle of the
pulse, which seems also to be the case in a number of pulsars. Also,
these fits back up the idea that very large pulses in Vela arrive
early, as pointed out by KD83.  Cairns et al. (2001) discuss the
implications for the emission mechanism of these log-normal
distributions.
\begin{table}
\begin{center}
\caption{\label{culmtab}
Log-normal fits to the flux density distribution}
\begin{tabular}{rrrr}
\hline
\hline
\multicolumn{1}{c}{Phase (ms)} & 
\multicolumn{1}{c}{Mean (Jy)} & 
\multicolumn{1}{c}{Sigma} \\
\hline
$-$0.575 &  3.5  & 0.60 \\
   0.319 & 35.0  & 0.22 \\
   1.213 & 13.0  & 0.25 \\
   2.107 & 12.5  & 0.19 \\
   3.001 &  3.25 & 0.27 \\
\hline
\end{tabular}
\end{center}
\end{table}

\subsection{The bump region}

\begin{figure*}

\centerline{\psfig{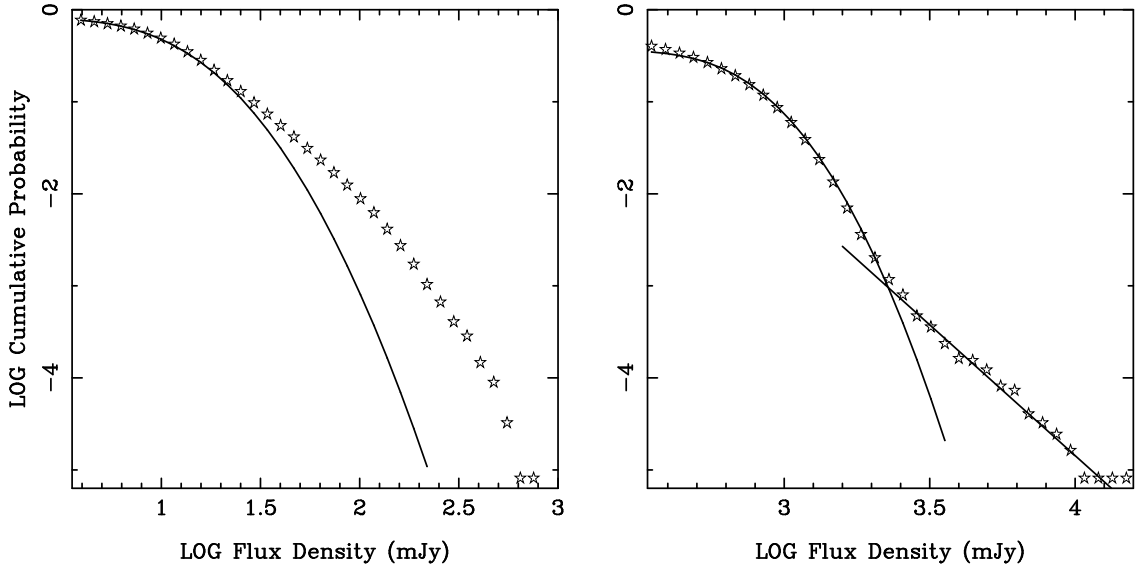}}

\caption{
\label{culmbump}\label{culmfiggiant}
Cumulative probability distribution of the integrated 
flux density under (a) the bump and (b) giant micro-pulse regions
at 2.3 GHz. 
The curved lines show the distribution expected from Gaussian
noise statistics which fits the data well at the low flux density
end but not at the high end. An additional power-law component
fits the data well in the right hand panel.}
\end{figure*}

We have attempted to isolate the bump component by choosing
a window 42 bins wide indicated by the arrows on the right hand
side shown 
Figure \ref{bumpfig}. We then computed the integrated flux density under
these 42 bins for each of the 123,000 single pulses.
Figure \ref{culmbump}a shows the cumulative
probability distribution along with the expected distribution
from the instrumental response (which dominates the non-bump signal at these
pulse phases). The presence of the bump component is clear in the
figure but the form of its distribution less so. A power-law distribution
overestimates the high flux probability. A log-normal distribution
also does not fit the data well unless a cut-off is imposed
at the low-flux end. This tends to suggest that the bump component
results from a separate emitting entity.

If the Vela pulsar consisted only of this component, its flux density
would be $\sim$14 mJy. The distribution of flux densities shows that
this component regularly has ``giant pulses'' with flux densities
exceeding $10\times$ the mean flux density
(Fig.~\ref{culmbump}a). There are 491 such pulses in the sample and so
a giant pulse occurs every 250 rotations.  These pulses contribute
17\% of the total flux density in these phase bins.  These are similar
properties to the giant pulses seen in the Crab
pulsar and it is tempting then also to relate the main pulse in Vela
to the so-called pre-cursor in the Crab. Such a picture is not
entirely convincing however, as the flux density distribution appears
not to be a power-law.  Moreover, there appears to be no coincidence
between the location of any of the radio emission and the high energy
emission from Vela, unlike the case of the Crab pulsar (see Moffett \&
Hankins 1999)\nocite{mh99}.  The geometric viewing angles in the two
cases are also rather different.  In Vela the line of sight cuts close
to the magnetic pole (Paper I) whereas in the Crab this angle is
rather large \cite{mh99}. Additionally,
Vela's high energy emission is complex with
the optical and soft X-ray emission occurring significantly later than
the hard X-ray and $\gamma$-ray emission \cite{shd99}.  Gouiffes
(1998)\nocite{gou98} has shown evidence for a component in the optical
which is at the same phase as the peak of the radio emission.  With
only $\sim$1 ms time-resolution it is not clear whether any optical
emission is associated with the radio bump component.

\subsection{Giant micro-pulses}


As previously discussed in Paper I, giant
micro-pulses of emission are present at phases just prior to the main
pulse.  These are very sporadic, and have significant phase jitter of
about 1 ms. We therefore chose a pulse window of about this size
located at early phases as shown in Figure \ref{bumpfig} to examine
the flux density distribution of these micro-pulses.  Figure
\ref{culmfiggiant}b shows the cumulative probability distribution of
the flux densities in these bins. It is evident that a power-law at high flux
densities is necessary.  The slope of the power law is $-2.85$,
similar to that seen in both the Crab pulsar and PSR B1937+21. The
occurrences of the giant micro-pulses are rather rare, only 60 pulses
out of 123,000 form the power-law tail in the figure.
Are the giant micro-pulses just an extension of the micro-structure
pattern to outside the main pulse window or are they more related to
the true giant pulses as seen in the Crab pulsar and PSR B1937+21?
Assuming that Eqn.~(\ref{scaling}) is also valid for very small periods
(cf.~discussion in Section~\ref{giant1}), one can
estimate an expected micro-structure time-scale for PSR B1937+21.
Interesting, the resulting time-scale of $\sim0.5$ $\mu$s is
coincident with the width measured for narrow (classical) giant pulses
by Kinkhabwala \& Thorsett (2000)\nocite{kt00} at 1.4 GHz and 2.3 GHz.
For the Crab pulsar the situation is more difficult. Its giant pulses
often appear to be severely scattered, but Sallmen et al.~(1999)
report that any intrinsic time-scale of the pulse is unresolved at 0.6
GHz ($\la 10$ $\mu$s), while at 4.9 GHz pulse components are typically
as short as 0.1 to 0.4 $\mu$s. This is much smaller than an
extrapolated time-scale of 14$\mu$s.

In summary, the observed energy distribution of the giant micro-pulses
bears similarities with that of known giant pulses (e.g.~Romani \&
Johnston 2001). The giant micro-pulses tend to be narrower than the
normal ones and, additionally, appear in a window outside that of the
normal average profile.  Hence, they may originate in the outer gap,
as has been argued for the classical giant pulses (Romani \& Johnston
2001\nocite{rj01}), unlike the bulk of the radio emission.


\section{Circular Polarization}

\label{polsection}

\begin{figure}

\centerline{\psfig{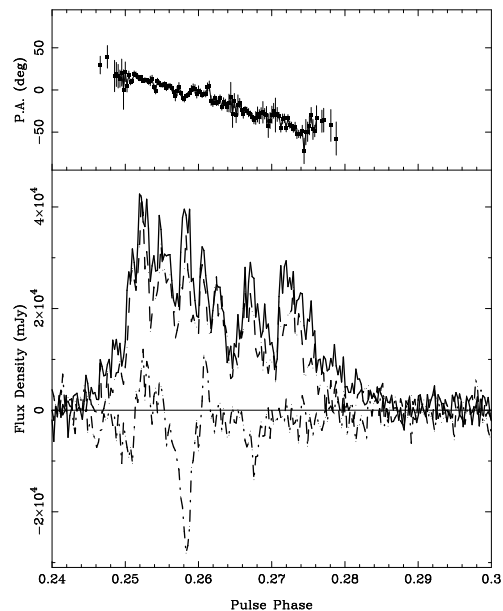}}

\caption{\label{polfig}
A single pulse in full polarization. The solid
line denotes total intensity, dashed line linear polarization and
dash-dotted line circular polarization. The top panel shows the
position angle. Note the high degree of correlation between the
total intensity and circular polarization and the sign reversal
of the circular polarization under some components. Also note that
the PA swing is not smooth but shows 'wiggles' with typical time
scales of a micro-pulse.}

\end{figure}

As is well known, Vela shows a very high fraction of
linear polarization at all pulse phases and for every pulse.
Apart from the bump region, there are no orthogonal mode
jumps at any pulse phase (KD83, Paper I). It is clear
therefore, that one mode completely dominates in the Vela pulsar
at all times; any other orthogonal mode, if present at all, must be
less than $\sim$5\% of the power of the dominant mode.
The circular polarization of the integrated profile increases
with increasing observing frequency and is about 15\% at
2.3 GHz with a predominantly negative sign throughout the pulse.
Individual pulses showing micro-structure appear to have
circular polarization which is highly correlated with the
total intensity. Occasionally the sign of circular polarization
can be positive under a given micro-structure feature. A typical
example of this is shown in Figure \ref{polfig}. Some micro-structure
features are found to be greater than 90\% circularly polarized
which poses challenges to emission theories.
It is important to re-iterate that change of the sign of the circular 
polarization is {\it not} accompanied by a orthogonal mode transition 
in the linear polarization (as is seen in other pulsars, see
e.g. Cordes \& Hankins 1977\nocite{ch77}).
This lack of mode transitions is not expected
in the model of McKinnon \& Stinebring (2000)\nocite{ms00}.

In order to study the relationship between total and circular
intensity (Stokes $I$ and $V$) for the micro-pulses in more detail, we
also performed the same micro-structure analysis on $V$ measured at 2.3
GHz.  In contrast to the procedure described in
Section~\ref{micromethod}, we used a fixed window of 128 phase bins
centred on the central pulse peak. Consistent with the total power
result summarized in Table \ref{resulttab}, we detect micro-structure
in $V$ in 83\% of all measured pulses. The determined time scale
$\tau_\mu$ of $87\pm43$ $\mu$s is somewhat smaller than that for $I$
at the same frequency but just consistent within the
uncertainties. 
We tentatively conclude
that the micro-structure properties seen for $V$ are consistent with
those seen in total power.
This is similar to the result found in other pulsars by Cordes
\& Hankins (1977)\nocite{ch77} and confirms the visual impression
(see Fig.~\ref{polfig}) 
that changes in the sign of the circular polarization
occur at the edges of the micro-structure rather than the middle.

\begin{figure}

\centerline{\psfig{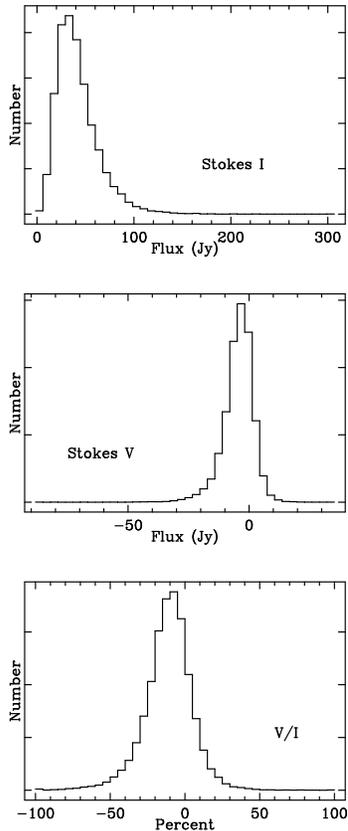}}

\caption{\label{vdist}
Distribution of (top) total intensity, (middle) circular polarization
and (bottom) percentage circular polarization  for 25000 pulses
at phase 0.32 ms.}

\end{figure}

We also examined the distribution of flux densities of $V$
in a given phase bin. Indeed, we find that the distribution is again 
log-normal with the same value of sigma as for $I$ in the
same phase bin. This is a remarkable result and implies that $V/I$ 
should be virtually constant from pulse to pulse in spite of large
variations in flux. The distribution of $V/I$ is a Gaussian with
a mean equal to the  $V$ of the integrated profile and a
sigma of 0.13. Figure \ref{vdist} shows these distributions.

In spite of the excellent signal-to-noise ratio of the data, the
distribution of $V/I$ is affected by noise. For example, when both
$V$ and $I$ are low then $V/I$ is essentially random and this
will cause a `pedestal' in the distribution. When $V$ is close to zero
this will cause an excess of counts near $V/I=0$ and the distribution
may appear peakier than Gaussian. To quantify this effect we
simulated the data in the following way.
First we picked a value for $I$ from a log-normal distribution
with a mean and sigma derived from the data as in Section 5.1.
We then picked a random value for $V$ from a Gaussian distribution
in $V/I$ with a mean equal to the integrated $V$ in that phase bin
and a sigma ($\sigma_v$) which is a free parameter. Following this,
Gaussian noise was added to both $I$ and $V$ (independently) and
the `observed' $V$, $I$, and $V/I$ were computed. This was repeated multiple
times and the output distributions could then be compared with the real data.

We find that in the centre of the pulse, when $\sigma_v =0$, the resultant
output distribution of $V/I$ has a width of 0.05. This is significantly
less than the true distribution. We therefore find we need an
intrinsic $\sigma_v$ of 0.11 to reproduce the observed data.
As we go towards the edges of the pulse profile, the observed value
of $\sigma_v$ becomes larger and larger. We find that virtually all
of this increase can be explained by the decrease in signal-to-noise
and that the underlying distribution of $\sigma_v$ is {\it constant} at 0.11.

\section{Discussion}

The origin of micro-structure is only one of the
many questions still to be answered when studying pulsar radio
emission. Micro-structure may be the product of either longitudinal
modulation of the radiation pattern and/or caused by radial modulation of
the emission or the creating plasma during the propagation in the
magnetosphere. In any case, these and previous results indicate that
micro-structure is an integral feature of pulsar emission that has to
be explained together with the observed fluctuations and polarization.

Popov et al.~(2001) found that the micro-pulse width of PSR B1133+16
scales with the width of the sub-pulse and the average component
width, indicating that micro-structure may be related to the sweep of
the beam.  
%
%
Interestingly, we find an obvious
period dependence of the micro-pulse width which is a indeed good argument
in favour of a temporal origin (i.e.~sweeping of beams of emission) of
micro-structure.  The exact $\tau_\mu-P$ relationship and any possible 
break-down of such at very small periods will be probed by observing
faster rotating pulsars and by studying Vela at an even higher frequency.
If a change of micro-pulse width with frequency is indeed confirmed,
we estimate that observations at 4.8 GHz should reveal a micro-structure 
width as predicted by Eqn.~(\ref{scaling}).

It is conceivable that the change in slope of the correlation of the
micro-pulses' peak flux density with their width shown in Figure
\ref{widthflux} could be an artifact due to the finite sampling time.
However, it could also indicate that there is an underlying population
of superimposed and blended, even narrower micro-pulses with widths of
typically 50 -- 100 $\mu$s.  Such narrow micro-pulses would be similar
in width to the giant micro-pulses although they would not have
similar peak flux densities, necessarily.  However, if there is a
population of narrow (normal) micro-pulses, it would be very difficult
to explain why we still observe quasi-periodic micro-pulses with a
preferred period -- despite the blending.  The preferred
quasi-periodicity is even constant for different frequencies, while
the observed average micro-pulse width is maybe not. Again, light can
be shed on these questions if simultaneous multi-frequency
observations of Vela micro-structure become available.

If confirmed,
the decrease in the observed average width of a micro-pulse is
stronger than but consistent with the general behaviour of the average
pulse profiles (or even sub-pulses) which is often attributed to a
Radius-to-Frequency mapping. If micro-structure represents micro-pulses
emitted at a particular altitude sweeping our line-of-sight, we would
indeed expect a smaller width at higher frequencies. The model of
``micro-beams'' creating the micro-pulses also finds support in the
observations of ``wiggles'' in the position angle (PA) swing of the
linearly polarized intensity.  Compared to the very smooth average PA
swing, that of individual pulses typically shows deviations from a
average slope on time scales consistent with an average micro-pulse
width, possibly relating these wiggles to PAs of separate micro-pulses
(see Fig.~\ref{polfig}).  The quasi-periodicities may then be caused
by a radial modulation of the emission. For instance, the spacing
between separate micro-pulses may be determined by a radial
distribution of emission patches, possibly arranged in a periodic way
during the creation of the emitting plasma or modulations during
propagation along the magnetic field lines.  If that is the case, we
would not necessarily expect a dependence of the quasi-periodicities,
$P_\mu$, on rotation period unless the mechanism is directly dependent
on the period value. Indeed, the data do not suggest any correlation
between pulsar spin period and the quasi-periods observed in
micro-pulses (see Table~\ref{previousobs}).

While we can conclude that {\em normal} micro-structure is an integral
part of pulsar radio emission, in agreement with the results of Lange
et al.~(1998) and Popov et al.~(2001), our study of the {\em giant}
micro-pulses shows that these are different. They appear occasionally
at pulse phases usually not covered by the normal pulses, have
typically large amplitudes, appear to be narrower than normal
micro-pulses, and also exhibit a power law in their cumulative
probability distribution. All these properties strongly suggest that
giant micro-pulses are a separate component of emission and more
closely related to the classical giant pulses.  The bump component is
yet another facet which appears to be a separate feature that may also
be related to giant pulses. Again, it appears only rarely and also
exhibits a energy distribution that is not compatible with a
log-normal distribution.  As pointed out, if Vela consisted of only
the bump component, it would regularly have classical giant pulses
with a flux density exceeding the mean flux density by an order of
magnitude. If classical giant pulses are related to high energy
emission as suggested by Romani \& Johnston (2001), and if giant
micro-pulses and the bump are in turn related to giant pulses, then the
absence of high-energy emission from these locations needs to be
explained.


\section{Summary}

We have detected micro-structure in about 80\% of all pulses,
independent of frequency.  Micro-pulses occurring in quasi-periodic
fashion show the same separation at 1.4 and 2.3 GHz but may appear
less frequently at the higher frequency. While there is a correlation
between micro-pulse width and corresponding peak flux density, the
width may decrease with frequency although this 
trend must be confirmed at higher frequencies. At 1.4 and 2.3 GHz, the
micro-pulse width is somewhat larger than expected from extrapolation
from other, slower rotating pulsars, although a period dependence is
clearly visible. Narrower, giant micro-pulses which occur at pulse phases
before the main pulse are indications of a possible relationship
between giant micro-pulses and classical giant pulses. This is
supported by our finding shown that the flux density distribution of
the giant micro-pulses is best fitted by a power-law. All micro-pulses
are highly linearly polarized and often show a large degree of
circular polarization, with a strong correlation between Stokes $V$
and $I$. 
We find that phase-resolved intensity distribution are best
fitted with log-normal statistics. We conclude that the ``bump''
component seems to be an extra component superposed on the main pulse

In summary, we find that Vela contains a mixture of emission
properties representing both ``classical'' properties of radio pulsars
and other radio features which are most likely related to high-energy
emission.  The Vela pulsar hence represents an ideal test case to study the
relationship between radio and high-energy emission in significant detail.

\section*{Acknowledgements}

This research was partly funded by grants from the Australian Research
Council. MK thanks the School of Physics of the University of Sydney
for their hospitality.  We are grateful to M.~Bailes for providing us
with the computer resources at the Swinburne Centre for Astrophysics
and Supercomputing to analyse our data. The Australia Telescope is
funded by the Commonwealth of Australia for operation as a National
Facility managed by the CSIRO.


\label{lastpage}

\end{document}